\documentstyle[aps,twocolumn,prl,epsf]{revtex}

\newcommand{\be}{\begin{equation}}
\newcommand{\ee}{\end{equation}}

\begin{document}

\twocolumn[
\hsize\textwidth\columnwidth\hsize\csname @twocolumnfalse\endcsname

\title{From Lorentz Force on Electron to  Magnus Force on
Vortex, \\  Role of Experiments }
\author{X.-M. Zhu, B. Sundqvist}
\address{Department of Experimental Physics, Ume\aa{\ }University,
    S-901 87, Ume\aa, Sweden}

\date{\today}
\maketitle

\widetext

\begin{abstract}

The vortex motion in a superfluid or a type II superconductor
is similar to the
electron motion in a magnetic field, because they both feel a
transverse force.
The vortex dynamics in a superconductor is a basic 
property of the superconductivity which remains controversial. 
It is also responsible for a large class of observed physical
phenomena.
We will examine this issue from the experimental point of view.
In particular, we will compare the
experiments which have set the stage to the Lorentz force
and the experiments influencing our
understanding of the Magnus force on vortices in superconductors.

\noindent PACS${\#}$s: 74.60.-w 
\end{abstract}
 
] 

\narrowtext

In hydrodynamics, we have learned a remarkable difference between
the Newtonian dynamics and the Eulerian dynamics. A particle obeying
Newton dynamics  accelerates along  the direction of applied force.
A vortex in a flow field, however, always has a tendency to
curve. If you push a vortex, it responds perpendicular
to your push, if the background flow is at rest. If a vortex is at
rest while there is a background flow, the vortex feels a
force perpendicular to the direction of the flow. In daily life, we encounter
numerous examples of this `curving' nature of vortices.

For an inviscous  hydrodynamic fluid,
 we have a well defined starting point,
the Euler equation,
\be
 \frac{\partial {\bf v}}{\partial t} + ({\bf v}\cdot\nabla ){\bf v}
= -\frac{\nabla P}{\rho} .
\ee
Here $\rho$ is the fluid mass density and $P$ is the pressure. The
above equation is applicable everywhere in space except at the
singular point of the vortex core.
The way we usually  derive the force on 
a vortex\cite{lamb} is to assume that
there is a trapped foreign object, for instance a disc,  in the
vortex core and study the force on this object.
The resulting Magnus force
is given by
\be
{\bf F} = \rho[ ( {\bf v}_0- {\bf v}_v )\times{\bf \kappa}].
\ee
Here $\kappa$ is the vorticity, ${\bf v}_v$ the velocity
of the vortex and ${\bf v}_0$ the velocity of the background 
flow. For a steady state motion, the Magnus force force 
balances the applied external force. If there is no external force,
the vortex always moves along the background flow.
The force on a vortex is similar to the force on an electron in 
a magnetic field. If we define a fictitious magnetic field
$\tilde{\bf B} =  \rho{\bf \kappa}$ and a fictitious electric field 
$\tilde{\bf E} = - \rho {\bf v}_0 \times{\bf \kappa}$, the force on a vortex
is identical to the force on an electron of a unit charge
${\bf F} = \tilde{\bf E} + {\bf v}\times \tilde{\bf B}$

In a neutral  superfluid He$^4$, we can employ the two fluid model
\cite{landau}.
The Euler equation for the superfluid component is rather similar
to that of hydrodynamic flow,
\be 
 \frac{\partial {\bf v}_s}{\partial t} + ({\bf v}_s\cdot\nabla ){\bf v}_s
= -\nabla\mu ,
\ee
where $\mu$ is the chemical potential. The Magnus force
also has a similar form,
\be
{\bf F} = \rho_s[ ({\bf v}_0 - {\bf v}_v)\times{\bf \kappa}].
\ee
Beside that the superfluid mass density $\rho_s$ replaces the
fluid mass density, there is also another important difference.
The vorticity is quantized in a superfluid so that the only way to
generate or annihilate a single vortex is to move it from(or to) the boundary.

For a charged superfluid, there are additional
terms due to the coupling to the electromagnetic field
in the Euler equation,
\be 
 \frac{\partial {\bf v}_s}{\partial t} + ({\bf v}_s\cdot\nabla ){\bf v}_s
= -\nabla\mu +\frac{e}{m} \left[
{\bf E} +\frac{1}{c}( {\bf v}_s \times {\bf H})\right].
\ee
Bearing in mind that the force is related to the
difference in electro-chemical potential difference,
we still obtain the same Magnus force as in the neutral
superfluid case\cite{nv}.

The type II superconductor is similar to a
fermionic superfluid. In principle, we should
be able to derive a hydrodynamic equation for the
the Cooper pairs, which can be regarded as bosons.
Such a hydrodynamic equation
has been written down  from the non-linear Schr\"odinger
equation for a superconductor in the clean limit 
at low temperatures\cite{nlse}.
In general, we should expect that a hydrodynamic equation
still exists, with a modified superfluid density and an effective
Cooper pair mass which can be determined experimentally.
Then we may ask why there is no consensus on the Magnus force 
in a superconductor\cite{re}
and where does the conflict of several different point of views 
originated.

The root of the disagreement is the experimental observation.
Unlike in the hydrodynamic fluid motion where we have plenty 
of examples to demonstrate the  Magnus force, in superconductors,
we do not have such obvious examples.
From the very beginning of the study of vortex dynamics in superconductors,
we have associated the understanding of the Hall effect to the  Magnus force.

Let us discuss in  more detail  the Hall effect in 
superconductors. For the time being
let us assume that the vortices in superconductors
feel the Magnus force from the hydrodynamic equation and examine
its consequences.
Under this assumption, the  equation of motion for a vortex  takes the
 form of the
Langevin equation similar to that of a
charged particle in the presence of a magnetic field:
\be
   m_v \ddot{\bf r} = q_v \frac{\rho_s}{2} h \; ( {\bf v}_s - \dot{\bf r} )
    \times \hat{z} - \eta \dot{\bf r} + {\bf F}_{pin} + {\bf f} \; ,
\ee
with an effective mass $m_v$,
a pinning force ${\bf F}_{pin}$, a vortex viscosity $\eta$, and a
fluctuating force ${\bf f}$. $q_v = \pm 1$ represent different vorticities.
The viscosity is related to the fluctuating
force by the usual fluctuation-dissipation theorem.

Now let us ignore the  vortex interaction as well as the pinnings.
Then we can solve the above equation to give 
\be
   \dot{\bf r} = \frac{ (\rho_s  h/2)^2}{ \eta^2 + (\rho_s  h/2)^2 } \;
                 {\bf v}_s
          + q_v \; \frac{(\rho_s  h/2) \eta }{ \eta^2 + (\rho_s  h/2)^2 } \;
                 {\bf v}_s \times\hat{z} \; .
\ee
According to the Josephson relation, the measured electric field ${\bf E}$
is given by 
\be
   {\bf E } = - q_v \frac{h}{2e} \; n \; {\bf v}_l \times \hat{z}  \; ,
\ee
and  it can be rewritten as
\be
      {\bf E } = - \frac{1}{c}   {\bf v}_l \times {\bf B}  \; .
\ee
We can calculate the longitudinal  and Hall  resistivity, $\rho_{xx}$ 
and  $\rho_{xy}$ from the above equations by
 $\rho_{xx} = E_x/J$ and $\rho_{yx} = E_y/J$. 
The friction can be estimated by assuming that the vortex
cores behave as normal electrons. Then we are ready to compare the
calculated $\rho_{xx}$ and  $\rho_{xy}$ with the experiments.

For the experiments which existed thirty years ago, the above results are
already qualitatively different from the experimentally measured ones. 
The Hall angle, defined by $tan^{-1}\theta =\rho_{xy}/  \rho_{xx}$
should be nearly $90$ degree for the estimated one while it was
only on the order of $0.5$ degree for the data existing
at that time. Naturally, the experimental facts led the theorists to
find a way out. In their work,
Bardeen and Stephen\cite{bs} argued that the Magnus force due to the
hydrodynamic flow of the Cooper pairs does not exist in
superconductor because the ionic background can contribute a
term to cancel the vortex velocity dependent part of the Magnus
force. Then the electrons at the nonsuperconducting vortex cores
can contribute to a  small Hall resistance in the same way as the normal 
electrons give rise to a Hall effect in an ordinary metal. The
question seemed to have obtained a satisfactory answer for  that time.

About thirty years later,
the Magnus force in a type II superconductor has generated a renewed
attention with the discovery of high-Tc superconductors 
especially after the observation of the so-called Hall anomaly.
It was found that the Hall conductance changes sign in the mixed
states in many  high-Tc superconductors and some conventional
ones. Apparently it can not be explained by the Bardeen-Stephen
model. Various attempts based on different ways to add  another small
contribution have been constructed. The contribution from the
particle-hole asymmetry and the contribution 
of the back-flow are such examples.
Overall speaking, the trend of the study of Magnus force 
on a superconductor was  motivated and led by the
experiments. The theories in this category have shown little  prediction 
power. 

\begin{figure}
   \def\epsfsize   #1#2{.5#1}
   \setlength {\unitlength}{1in}
    \vskip 1.0cm
   \begin {picture}(3,2.7)(-0.2,-0.1)
      \epsffile {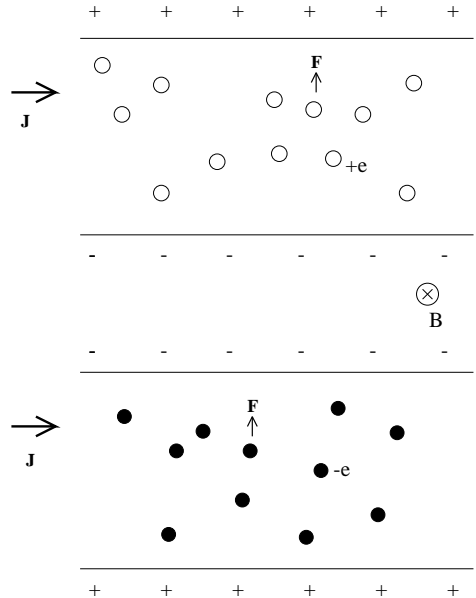}
   \end {picture}
\caption{ When an electric current is passed through a metal bar
in a magnetic field, the
Hall voltage depends on the carrier type. The force this metal bar
feels only depends on the total electric current carried through, not on how
it is carried through.
The upper graph shows the hole carriers and the lower graph shows
the electron carriers. }
\end{figure}

The situation concerning the Lorentz force on an electron is
very different. It might be helpful to examine the history on this
undoubtedly successful sub-field of electron theory and see if we can 
learn anything from it. We have already shown that the vortex dynamics
is similar to the electron dynamics in a magnetic field.  What is the
role that the  Hall effect played in determining the Lorentz force
and establishing the theory of electrons? We have to say little, if there
is anything at all. Then what does this simple fact tell us?

Let us recall how  we first introduce the 
Lorentz force in a text book? Take for
example {\it The Feynman Lectures on Physics}\cite{feynman}.
 It was introduced
with two experiments. The first one measures the force on a wire carrying an
electric current near a bar magnet, the second one the force 
between two wires carrying electric current simultaneously. 
In other words, the Lorentz force was introduced by genuine force
measurement experiments. It was by these types of force measurements that
we started to realize there is an interaction between moving charges and
the magnetic field. What would have happened if we did not have these
type of experiments but only had the Hall effect? We probably would have taken
a long detour toward understanding the Lorentz force and developing the theory
of electrons.

There is a reason to believe the Hall effect in a type II superconductor
has led the theorists to such a detour. The Hall effect in a superconductor
is in many ways similar to that of a normal metal. In a normal metal,
the electrons interact with the background lattice potential so that we
have Bloch bands. They also interact with each other through the 
Pauli principle.
As a result, the bandstructure determines the carrier types.
The Hall coefficient can be positive or negative depending on the
details. Although this is well known today, it was understood decades after
the Lorentz force itself. If we did not know the Lorentz force beforehand,
it would have been very difficult trying to understand  the Lorentz force
and the Hall effect at the same time. The vortices in a superconductor
also feel background potentials. Pinnings from inhomogeneity  
inevitably exist. In addition, there is also an intrinsic pinning due
to the interaction between
the vortex and the lattice background. This coupling exists even when there are
no defects in the sample. The vortices also interact with each other and
they form a lattice at lower temperatures.
When they form a lattice, the sliding
of the whole lattice is only one of the modes of motion. Vortices can
move by defect motion too. It has been pointed out that defect 
configurations act like different carrier types.
Even when the temperature is higher than the melting temperature
of the vortex lattice, the vortex-vortex
interactions still cannot be ignored.
We know perfectly well that many transport properties of a liquid
are considerably influenced by the interactions. 

In fact, without any arm-twisting modification of the basic superfluid
hydrodynamics which governs the vortex motion, the Hall effect
in a superconductor can also be explained in a similar way
as the Hall effect in a normal metal. In such a framework, 
the Magnus force, like the Lorentz
force is a basic property and is not sample dependent\cite{at}.
The vortex motion, however, is  complicated by all kinds of possible
defect motions. At low temperatures, the vacancy motion may
dominate and give rise to the Hall anomaly\cite{vacancy}. 

Can we put the above framework into test experimentally?
There is a  difference between vortices and
electrons concerning measurement.
It is possible to study a single electron motion in vacuum so that
it is free from any of the complications we face in a Hall effect
measurement. However, for the superconductor, an individual
vortex cannot penetrate through the sample.
As long as the magnetic
field is above $H_{c1}$, the vortices always form a lattice so that we
have to consider vortex-vortex interaction.
The situation is not hopeless, however. Still, much can be learned from
the experiments we used to introduce the Lorentz force.
In our description of those experiments, we never considered the lattice
background potential or the Fermi statistics. It was not even known at
the time this type of experiments was first performed. Apparently
those experiments are independent of carrier types. 
We will demonstrate more in fig.1. If we pass a current through a
metal bar in a magnetic field, depending on the bandstructure of
this metal,   the
Hall effect may have  different signs. However,
the force on this metal bar is given  by 
the total Lorentz force ${\bf J}\times{\bf B}$ 
and is independent of the details on how the electrons
are carried from one end of the bar to the other.

This simple experimental construction tells us an important way to
construct an experiment to measure the Magnus force on the vortex.
If we can  follow the same principle
to construct a direct force measurement when the vortices plays the
role of electrons, we should be able to determine the
Magnus force even we do not know how the vortices actually move, by
defect motion, plastic flow or other ways.

\begin{figure}
   \def\epsfsize   #1#2{.5#1}
   \setlength {\unitlength}{1in}
    \vskip 1.0cm
   \begin {picture}(3,2.7)(-0.2,-0.1)
      \epsffile {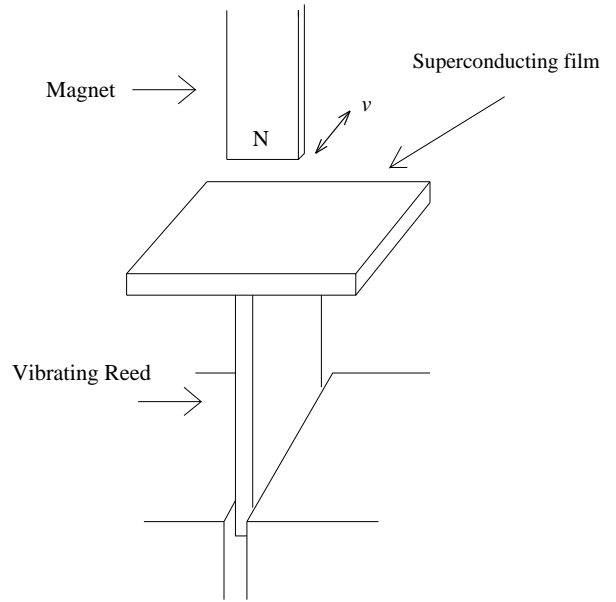}
   \end {picture}
\caption{Schematic drawing of the experimental setup used to
observe the Magnus force on the vortices in a superconductor.}
\end{figure}

In principle we can construct a experiment almost exactly as the
first experiment shown in the textbook \cite{feynman}. We can
attach two wires to a superconductor,
place it in a magnetic field and pass a current through. 
Then we can measure the force on the superconductor.
We also need to determine the direction and the magnitude of the
vortex current. This is not a problem if we can measure the
longitudinal and Hall resistance at the same time. There is only 
one drawback that all the high Tc superconductors are not
wire-friendly. It is not easy to attach and manage so many wires to
a high Tc superconductor and perform a force measurement at the same
time.

A direct force experiment was carried out
recently with a slightly different design.
The vortices were driven into motion by a small magnet
which was vibrating 
above a superconducting film. 
When  the magnet is moving at a low speed,  we can neglect the
vortex mass and assume the  total force on each
vortex  to balance. On each vortex, there is a 
 force from the moving magnet, a  transverse force 
acting on the vortex from the superconductor (the Magnus force),
a  pinning  force and a friction force from the underlying lattice 
and  the interaction force from other vortices. Now let us examine
the total force summed over all vortices. 
The  total force of vortex interactions vanishes. 
In the direction  parallel to the motion of the magnet,
the total force from the moving magnet to the
vortices will balance the total pinning   and the  
total friction from the superconductor.
In the transverse direction, the total force from the moving magnet to the
vortices  will balance
the total transverse force on vortices 
from the superconductor, i.e. the total velocity dependent part 
of the Magnus force.
Thus in the transverse direction, a reaction force to the
total Magnus force  which is  coming from the
moving magnet to the vortices will be passed entirely to
the superconductor. This is exactly the force
the superconducting film feels. To measure this force on the film,
the film is mounted on a vibrating reed and the frequency of the
vibration of the magnet is tuned to  sweep through the resonance
frequency of the vibrating reed. To maximize the
signal, the direction of the vibration of
the reed is adjusted to be perpendicular to the motion of the
magnet.

Indeed this force measurement has provided us with something valuable.
The Magnus force was found to have the same sign and
the order of magnitude as predicted from a hydrodynamic equation.
Although more experiments are still needed to
complete this subject, at least we  started to have  a new direction to
design  and to understand our experiments. The Hall effect is a
very interesting subject by itself. However, it did not help  the
development of the theory of electrons and it most likely  will not
help developing the theory of vortex dynamics in superconductors.
We can put the Hall effect aside when we try to derive the
equation of motion for the vortex.
We only need to study the theory  from a theoretical point of view.
Hopefully, something new can come out of it.

This work was financially supported by the Swedish Natural Science Research
Council (NFR) and the Swedish Research Council for the Engineering Sciences
(TFR).

\end{document}